\documentstyle[preprint,prb,aps]{revtex}

\begin{document}

\title{Effective electronic response of a system of metallic cylinders}
\author{J. M. Pitarke$^1$, F. J. Garc\'\i a-Vidal$^2$, and
J. B. Pendry$^3$} 
\address{$^1$Materia Kondentsatuaren Fisika Saila, Zientzi
Fakultatea, Euskal Herriko  Unibertsitatea,\\ 644 Posta kutxatila, 48080 Bilbo,
Basque Country, Spain\\
$^2$Departamento de F\'\i sica Te\'orica de la Materia Condensada,
Facultad de Ciencias,\\ Universidad Aut\'onoma de Madrid, 28049 Madrid, Spain\\
$^3$Condensed Matter Theory Group, The Blacket Laboratory, Imperial College,\\ 
London SW7 2BZ, United Kingdom}
\date{March 9 1998}
\maketitle

\begin{abstract}

The electronic response of a composite consisting of aligned metallic 
cylinders in vacuum is investigated, on the basis of photonic band 
structure calculations. The effective long-wavelength dielectric response 
function is computed, as a function of the filling fraction. A spectral 
representation of the effective response is considered, and the surface 
mode strengths and positions are analyzed. The range of validity of a 
Maxwell-Garnett-like approach is discussed, and the impact of our results 
on absorption spectra and electron energy-loss phenomena is addressed.

\end{abstract}


\newpage

\section{Introduction}

Both theoretical and experimental investigations of the optical properties 
of composite materials have been of basic importance over the years
\cite{Bergman1}. In particular, effective medium theories, in which the electromagnetic
response of the composite is described in terms of an effective dielectric 
function $\epsilon_{\rm eff}$, have been a useful tool for the interpretation of
absorption spectra\cite{Landauer}. Mean-field  theories of the effective response
function have also been proved to be  useful to analyze electron energy-loss experiments
in which swift electrons penetrate composites at random distances from the
particles\cite{Echenique}$^-$\cite{Pitarke1}.

In the long-wavelength limit, the effective transverse dielectric function for 
a system of spherical particles was first derived by Maxwell-Garnett (MG)\cite{MG},  
within a mean-field approximation valid for small values of the volume
occupied by the spheres. A generalization of the MG theory to finite 
wavelength, taking account of the spatial nonlocality in the response, 
has been recently presented by Barrera and Fuchs\cite{Barrera}. On the other hand, many 
attempts have been made to account, at large filling fractions, for higher 
multipole interactions, which are absent in the MG theory\cite{Wood}$^-$\cite{Bergman2}.
Recently, exact numerical calculations for periodic dielectric structures have been 
performed\cite{Datta}, based in the long-wavelength limit of photonic band structure 
calculations, and the MG theory has been shown to offer a good  approximation  in the
low-filling-ratio regime. Finally, the discovery of  tubular fullerenes\cite{Ijima}, a
few years ago, has opened a new focus of interest for  the application of these theories
to the case of cylindrical interfaces\cite{Bursill,FJ}.

In this paper we evaluate, on the basis of photonic band structure 
calculations, the long-wavelength limit of the effective transverse dielectric 
function of a composite made up of long metallic cylinders embedded in an 
otherwise homogeneous medium. We first demonstrate that the composite can be replaced by
an effective homogeneous medium, and then we determine the transverse dielectric
function of that medium from the electromagnetic modes supported by such a homogeneous
system. We consider a spectral  representation of the effective dielectric function, we
present an  evaluation of the parameters involved, and we discuss the range of validity 
of a MG-like approach appropriate for cylindrical particles. We also  address the impact
of our results on absorption spectra and electron  energy-loss phenomena.

\section{Theory}

It was shown by Bergman\cite{Bergman3} and Milton\cite{Milton} that in
the long-wavelength limit the effective dielectric function of any macroscopically
homogeneous two-component system of dielectric constants $\epsilon$ and $\epsilon_0$
and volume fractions $f$ and $1-f$, respectively, can be expressed as a sum of simple
poles which depend only on the microgeometry of the composite material and not on the
dielectric functions of the components:
\begin{equation}     
\epsilon_{\rm eff}=\epsilon_0\left[1-f\sum_\nu{B_\nu\over u-m_\nu}\right],
\end{equation}
where $u$ is the spectral variable
\begin{equation}
u=\left[1-\epsilon/\epsilon_0\right]^{-1},
\end{equation}
$m_\nu$ are depolarization factors,
and $B_\nu$ are the strengths of the corresponding normal modes, which all add up to
unity:
\begin{equation}
\sum_\nu B_\nu=1.
\end{equation}

Similarly\cite{note1},
\begin{equation}     
\epsilon^{-1}_{\rm eff}=\epsilon_0^{-1}\left[1-f\sum_\nu{C_\nu\over u-n_\nu}\right]
\end{equation}
and
\begin{equation}
\sum_\nu C_\nu=1.
\end{equation}
In particular, if there is only one mode with a strength different from zero, one
finds:
\begin{equation}
\epsilon_{\rm eff}(\omega)=\epsilon_0\left[1-f{1\over u-m_1}\right]
\end{equation}
and
\begin{equation}
\epsilon_{\rm eff}^{-1}(\omega)=\epsilon_0^{-1}\left[1+f{1\over u-n_1}\right].
\end{equation}

We consider a binary composite with a volume fraction $1-f$ of insulator, with
real dielectric constant $\epsilon_0$, and $f$ of a periodic  system of long metallic
cylinders of diameter $d$ arranged in a square array  with lattice constant $a=x\,d$
(See Fig. 1). The diameter of the cylinders  is taken to be small in comparison with the
wavelength of the  electromagnetic excitation and large enough that a macroscopic
dielectric  function $\epsilon(\omega)$ is ascribable to the cylinders. For 
simplicity, the magnetic permeabilities are assumed to be equal to unity in 
both media.
 
Now we take an electromagnetic (EM) wave normally incident on the structure, 
so that $k_x=k_y=0$. For this propagation direction there are two different 
values of $\epsilon_{\rm eff}(\omega)$ corresponding to different polarizations. In the 
case of EM waves polarized along the cylinders ($s$ polarization), the 
presence of the interfaces does not modify the electric field, and one 
easily finds\cite{note2} that the response of the composite is equivalent to that of a 
homogeneous medium having the effective transverse dielectric function of Eqs. (6) and
(7) with $m_1=0$ and $n_1=f$.

For EM waves polarized normal to the cylinders ($p$ polarization), we 
first consider the case of a single cylinder embedded in an otherwise 
homogeneous medium. An elementary analysis of this classical problem shows 
that the electric field in the interior of the cylinder is\cite{Bohren}
\begin{equation}
{\bf E}_{\rm in}={u\over u-1/2}{\bf E}_{\rm out},
\end{equation}
where ${\bf E}_{\rm out}$ represents the electric field intensity in the medium
outside the cylinder. The effective dielectric function of isolated cylinders is then
found to be given, again, by Eqs. (6) and (7), the depolarization factors now being
$m_1=n_1=1/2$. In the case of isolated particles only the dipole surface mode enters Eqs.
(1) and (4) for this polarization.

The interaction between the particles in a composite has been considered, over the
years, within the well-known MG approximation. The basic assumption of this approach is
that the average electric  field within a particle located in a system of identical
particles is related  to the average field in the medium outside as in the case of  a
single isolated particle, thus only dipole interactions being taken into account. Hence,
after replacing, for $p$ polarization,
${\bf E}_{\rm out}$ of Eq. (8) by
\begin{equation}
{\bf E}_{\rm out}={<{\bf E}>-f{\bf E}_{\rm in}\over 1-f},
\end{equation}
$<{\bf E}>$ being the macroscopic effective electric field in the composite, the
MG result for cylindrical inclusions is found, which can be represented by the use of
Eqs. (6) and (7)  with
\begin{equation}
m_1={1\over 2}(1-f)
\end{equation}
and
\begin{equation}
n_1={1\over 2}(1+f).
\end{equation}
This is a special case of the generalized MG dielectric function derived in
Ref.\onlinecite{Bohren} for ellipsoidal inclusions, which in the case of aligned
spheroids distributed at random has been extended beyond the MG
approximation\cite{Barrera2}. Higher multipole interactions are neglected in these
approaches.

In order to compute, with full inclusion of higher multipoles, the effective dielectric 
function of our periodic system, we first follow Ref.\onlinecite{Pendry1} to calculate
the photonic band structure. We fix the frequency, the local dielectric function
inside the cylinders being therefore specified, and we discretize the Maxwell equations
within the unit cell. The discretized Maxwell equations provide a relationship between
the electromagnetic fields on either side of the unit cell, i.e., $z$ and $z+a$ (see
Fig. 1), and we exploit this to calculate, on applying Bloch's theorem, the eigenvalues
of the so-called transfer matrix, which will give the band structure of the system.

Homogeneous media are well known to support only two transverse modes. On the other
hand, looking at the imaginary part of the eigenvalues of the system, we observe that
over the entire range of frequencies there are, for each polarization, two degenerate EM
Bloch waves with vectors $k$ and $-k$ for which
$k(\omega)$ roughly follows the dispersion relation of free light and which are
dominant in the sense that they present the smallest attenuation (smallest imaginary
part of $k$). Thus, we conclude
that the composite can be replaced by an effective homogeneous medium, and we use the
dispersion relation of these Bloch waves to compute the effective transverse dielectric
function from
\begin{equation}
\epsilon_{\rm eff}(\omega)={k^2c^2\over\omega^2},
\end{equation}   
where $c$ is the speed of light.

\section{Results and discussion}

We have taken the insulating component of the composite to have a real
dielectric constant of $\epsilon_0=1$, and we have modeled the dielectric properties of
the metal inside the cylinders  by a simple free-electron (Drude) dielectric function:
\begin{equation}
\epsilon(\omega)=1-{\omega_p^2\over\omega(\omega+{\rm i}\gamma)},
\end{equation}
where $\omega_p$ is the bulk plasma frequency of the metal, and $\gamma$ 
represents an inverse electron relaxation time. We have used 
the plasma frequency of the conduction electrons in Aluminum,
$\hbar\omega_p=15.8{\rm eV}$, and $\gamma=1.4{\rm eV}$.

For $s$ polarized electromagnetic waves, our numerical results for the 
effective dielectric function, as obtained from Eq. (12), accurately reproduce the
exact results of Eq. (6) and (7) with $m=0$ and $n=f$, and this represents, therefore, a
good check of our scheme.

In the case of $p$ polarized electromagnetic excitations, great care was 
exercised to ensure that our calculations are insensitive to the precise 
value of the number of mesh points in the unit cell\cite{FJ0}. The results presented 
below have been found to be well-converged for all filling fractions, and 
they all correspond to a sampling mesh of $180\times 1\times 180$\cite{note3}.

The optical absorption is directly given by the imaginary part of the effective
dielectric function, ${\rm Im}\left[\epsilon_{\rm
eff}(\omega)\right]$. Thus, for $p$ polarization the absorption spectra of isolated
cylinders ($f\to 0$) exhibit a strong maximum, absent in the bulk metal, at the cylinder
dipole resonance ($\epsilon(\omega)+\epsilon_0=0$), i.e., at
$\omega_1=\sqrt{m_1}\omega_p$ with $m_1=1/2$. At higher concentrations of metal, ${\rm
Im}\left[\epsilon_{\rm eff}(\omega)\right]$ still shows a dipole resonance at
$\omega_1=\sqrt{m_1(f)}\omega_p$ ($m_1(f)\neq 1/2$), and a band of multipole resonances
at $\omega_\nu=\sqrt{m_\nu(f)}\omega_p$ ($\nu=2,...$ ; $m_\nu>m_1$) is also formed
broadened by electromagnetic interactions among individual cylinders. This is
illustrated in Fig. 2, where our first principles calculations of ${\rm
Im}\left[\epsilon_{\rm eff}(\omega)\right]/f$ are exhibited, as a function of the
frequency $\omega$, for various values of the ratio $x$
between the lattice constant and the diameter of the cylinders. For $x\ge 2$ the
effective dielectric function is well described by the MG approximation, and both our
computed result and the MG approximation almost coincide for
$x=6$ with the isolated cylinder result, represented in Fig. 2 by a thin solid line. In
Fig. 3 our results are compared, for two different filling fractions, with the MG
approximation, showing that together with the appearance of multipole resonances the
depolarization factors $m_1(f)$ are, for $x<2$, smaller than the MG depolarization
factor of Eq. (10).
  
In the case of a broad beam of charged particles penetrating the composite, the
probability per unit path length, per unit energy, for the projectile to transfer energy
$\hbar\omega$ to the medium is\cite{Pitarke2}
\begin{equation}
P_{\omega}={2e^2\over\hbar^2 v^2}F(\omega),
\end{equation}
where
\begin{equation}
F(\omega)={1\over\pi}\int_0^{Q_c}{\rm d}Q{Q\over
Q^2+q_v^2}{\rm Im}\left[-\epsilon_{eff}^{-1}({\bf q},\omega)\right],
\end{equation}
${\bf v}$ and $\epsilon_{eff}^{-1}({\bf q},\omega)$ being the velocity of
the incident particles and the effective inverse longitudinal dielectric function of the
composite, respectively. $\hbar{\bf Q}$ represents the component of the momentum
transfer, $\hbar{\bf q}$, located in a plane perpendicular to the incident beam
direction,
$Q_c$ is determined by $q_c^2=Q_c^2+q_v^2$, $q_c$ representing the largest wave vector
for which plasmons are well-defined excitations\cite{note4}, and
\begin{equation}
q_v={\omega\over v}.
\end{equation}

In typical electron energy-loss experiments swift electrons of about $100{\rm keV}$
penetrate the composite, and the main contribution to the integral of Eq. (15) comes,
therefore, from very small values of the momentum transfer. Also, for small values
of the adimensional parameter $qa$ ($qa<1$), $a$ being the radius of the cylinder, the
effective dielectric function of isolated cylinders has been proved to be rather
insensitive to the introduction of a finite wave vector\cite{Pitarke1,Pitarke2}, and the
energy-loss spectra of swift electrons is therefore expected to be, for small values of
the radius of the cylinders ($a<1{\rm nm}$)\cite{note6}, well described by the $q\to 0$
limit of the imaginary part of the effective inverse dielectric function\cite{note5}.

Eq. (7) with $n=0$ (dilute limit of $n=f$) and $n=1/2$ exactly coincides, in the
long-wavelength limit, with the result one obtains for the effective inverse
longitudinal dielectric function of isolated cylinders when the momentum transfer is
parallel to the axis and when it is located in a plane perpendicular to the axis of the
cylinder, respectively\cite{Pitarke2}. In this limit ($q\to 0$) bulk plasmons are not
excited, as a consequence of the so-called Bregenzung effect, and the pole in Eq. (7) at
$\epsilon(\omega)+\epsilon_0=0$, i.e., $\omega_1=\sqrt{m_1}\omega_p$ with $m_1=1/2$,
determines, for $p$ polarization, the frequency of the so-called surface plasmon
resonance at which electron energy-loss occurs. 

Fig. 4 shows, as a function of the frequency, our calculations of ${\rm
Im}[-\epsilon^{-1}_{\rm eff}(\omega)]/f$ obtained for various filling fractions by
varying the ratio $x$ between the lattice constant and the diameter of the cylinders. In
a homogeneous metal the so-called energy-loss function, ${\rm
Im}[-\epsilon^{-1}_{\rm eff}(\omega)]$, shows a single peak at the bulk
plasmon resonance, and this peak persists in the composite for
$x\le 1$, i.e., for filling fractions of metal for which the insulator no longer forms a
connected medium. Besides this peak, there is also a band in ${\rm
Im}[-\epsilon^{-1}_{\rm eff}(\omega)]$ from multipole contributions to the effective
response at $\omega_\nu=\sqrt{n_\nu(f)}\omega_p$ ($\nu=2,...$ ; $n_\nu<n_1$). As in the
case of
${\rm Im}\left[\epsilon_{\rm eff}(\omega)\right]$, the energy-loss function is well
described by the MG approximation in the low-filling-ratio regime, and our results
begin to deviate from those obtained within the MG approximation at
$x\approx2$ (see Fig. 5). We also find that the depolarization factors $n_1(f)$ are
larger than the MG depolarization factor of Eq. (11) for all volume fractions of metal
for which the MG theory does not reproduce our results.

Our numerical calculations of the effective dielectric function, as obtained from Eq.
(12), lead us to the conclusion that the depolarization factors $m_1$ and $n_1$ of Eqs.
(1) and (4), corresponding to dipolar modes, satisfy the following relation:
\begin{equation}
n_1=1-(D-1)m_1,
\end{equation}
where $D$ represents the dimensionality of the inclusions, i.e., $D=3$ in the case
of spherical inclusions and $D=2$ in the case of long circular cylinders. If all
multipolar modes are neglected, then $B_1=C_1=1$ in Eqs. (1) and (4) and a combination of
these equations with Eq. (17) results, for $D=2$, in the spectral representations
of Eqs. (6) and (7) with the depolarization factors $m_1$ and $n_1$ given by Eqs. (10)
and (11), i.e., the MG approximation. On the other hand, as long as multipolar
modes give non-negligible contributions to the spectral representation of the effective
response, i.e., $B_\nu\neq 0$ and $C_\nu\neq 0$ ($\nu=2,...$), the strengths of the
dipolar modes, $B_1$ and $C_1$, become smaller than unity (see Eqs. (3) and (5)), and a
combination of Eqs. (1) and (4) with Eq. (17) leads us to the conclusion that the
depolarization factors $n_1$ and $m_1$ necessarily deviate from their MG counterparts.
That a non-vanishing contribution from multipolar modes appears together with a
deviation of the dipolar mode locations with respect to their MG counterparts is
apparent from Figs. 3 and 5.

One of the main advantages of the spectral representation of Eqs. (1) and (4) is
that the depolarization factors and the strengths of
the corresponding normal modes depend only on the microgeometry of the composite
material and not on the dielectric function of the components. On the other hand, the
normal mode positions are determined by $m_\nu$ and $n_\nu$, and
we have thus represented in Fig. 6, in the case of $p$ polarization, universal curves
for the actual dipolar mode depolarization factors and strengths versus the filling
fraction. It is obvious from this figure that the trend with increasing filling fraction
is for the dipolar peaks in ${\rm Im}\left[\epsilon_{\rm eff}(\omega)\right]$ / ${\rm
Im}\left[-\epsilon^{-1}_{\rm eff}(\omega)\right]$ to move from the cylinder dipole mode
($m_1=n_1=1/2$) to lower / higher energies. Notice that the MG results, also plotted in
this figure, agree with our numerical results for $f<0.2$ ($x>2$). Nevertheless, at
higher filling fractions the actual depolarization factors approach more rapidly, as
$f$ is increased, the depolarization factors appropriate to the homogeneous metal
($m_1=0$ and $n_1=1$). In particular, for $x<1$ ($f>\pi/4$) $n_1=1$, indicating that
losses due to the bulk plasmon are expected when the metal forms a connected medium.

Associated to the deviation of the actual dipolar mode positions from the MG results
is the appearance of non-negligible multipolar mode strengths, $B_\nu$ ($\nu=2,...$) and
$C_\nu$ ($\nu=2,...$), and, accordingly (see Eqs. (3) and (5)), a reduction in the
dipolar mode strengths $B_1$ and $C_1$. These mode strengths determine the
dipolar peak heights in ${\rm Im}\left[\epsilon_{\rm eff}(\omega)\right]$ and
${\rm Im}\left[-\epsilon^{-1}_{\rm eff}(\omega)\right]$, represented in the inset of
Fig. 6 as a function of $f$. For $B_1=C_1=1$, all multipolar mode strengths being
neglected, the dipolar peak heights in ${\rm Im}\left[\epsilon_{\rm eff}(\omega)\right]$
and
${\rm Im}\left[-\epsilon^{-1}_{\rm eff}(\omega)\right]$ are found to be
$f\,H/\sqrt{m_1}$ and
$f\,H/\sqrt{n_1}$, respectively, where $H$ represents the peak height in the bulk
energy-loss function (in the case of the Drude dielectric function of Eq. (13),
$H=\omega_p/\gamma$). The actual dipolar peak heights divided by $f\,H$ are represented
by stars and dots, together with the $B_1=C_1=1$ approximation, i.e., $1/\sqrt{m_1}$ and
$1/\sqrt{n_1}$, showing, therefore, the magnitude of the actual dipolar mode strengths
involved in the spectral representations of Eqs. (1) and (4).

In summary, we have evaluated, on the basis of photonic band structure
calculations, the effective long-wavelength dielectric response function of a system of
long metallic cylinders. We have considered a spectral representation of both direct and
inverse effective dielectric functions, of interest in the interpretation of absorption
spectra and electron energy-loss experiments, respectively. We have analyzed the surface
mode strengths and positions, and the effect of the electromagnetic interactions between
the cylinders has been investigated. We have concluded that MG results are good as long
as the distance between the axis of neighbouring cylinders is larger than twice the
diameter of the cylinders. This is in contrast with the results we have also obtained
for long dielectric cylinders. In this case we have found the MG results to be good for
almost all the filling ratios for which the cylinders are not touching, as happens in the
case of a periodic array of dielectric spheres\cite{Datta}.

\acknowledgments

J. M. P. acknowledges partial support by the University of the Basque
Country and the Basque Unibertsitate eta Ikerketa Saila 
under contracts UPV063.310EA056/96 and GV063.310-0017/95, respectively, and, also, by
the British Council.

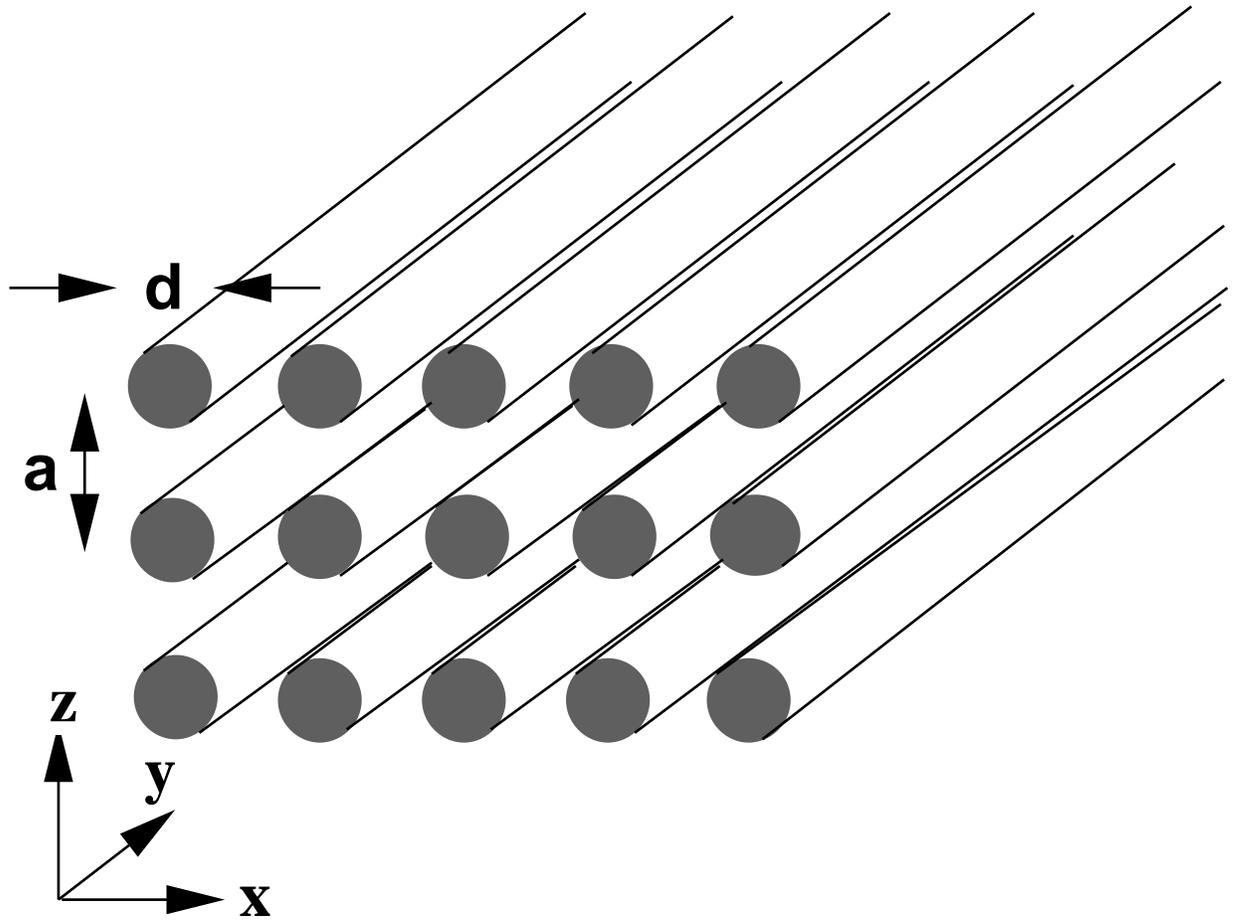
\begin{figure}
\caption{Periodic 
array of metallic cylinders of diameter $d$ arranged in a square array 
with lattice constant $a$. The cylinders are infinitely long in the $y$-direction.}
\end{figure}

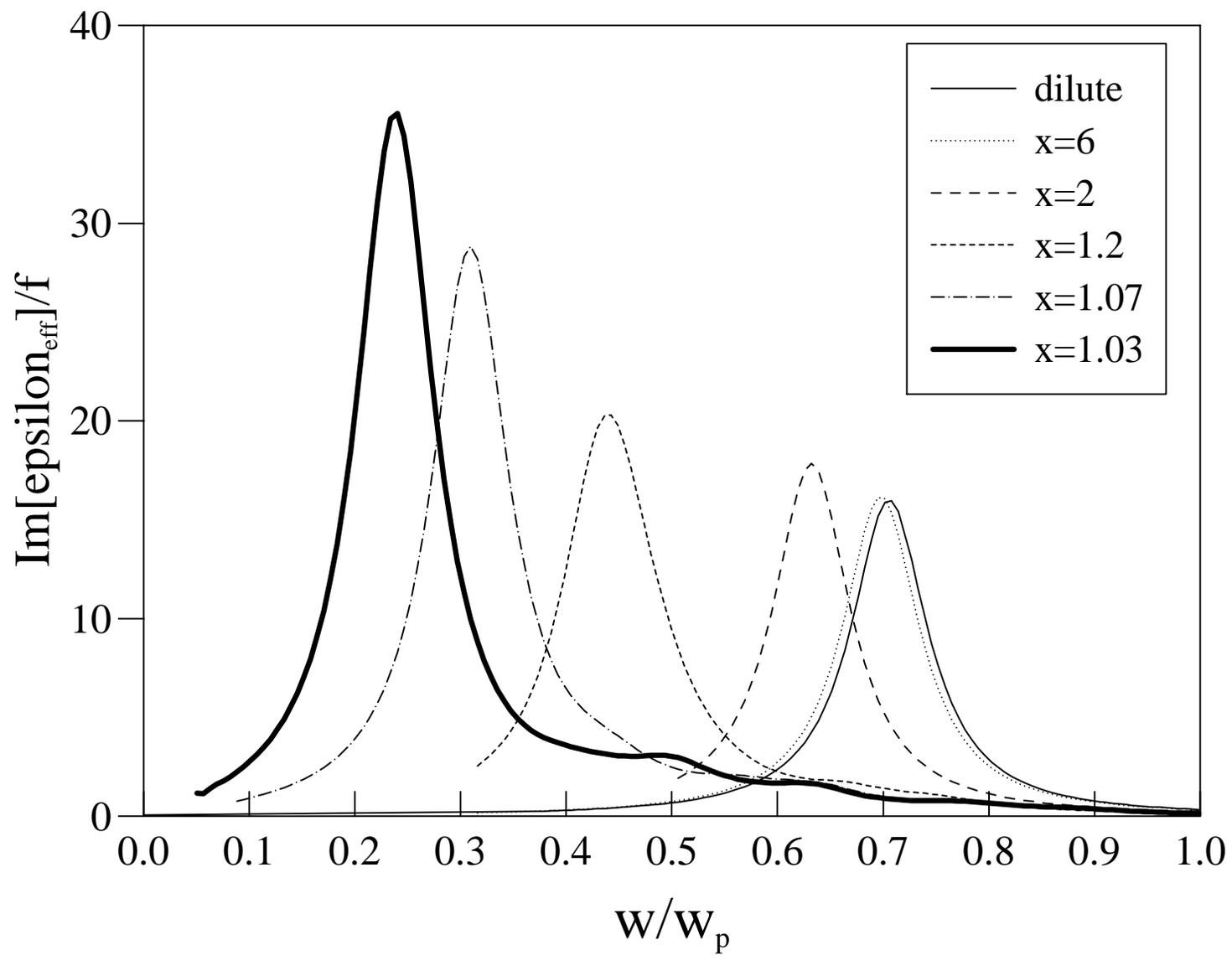
\begin{figure}
\caption{The imaginary part of the effective long-wavelength dielectric function of the
periodic system described in Fig. 1, for $p$ polarized electromagnetic excitations. Full
thin line: the isolated cylinder result ($f\to 0$). Dotted, long-dashed, short-dashed,
dashed-dotted and thick full lines: our numerical results for volume filling fractions
of $2.2\%$, $19.6\%$, $54.5\%$, $68.6\%$ and $74.0\%$, respectively.}
\end{figure}

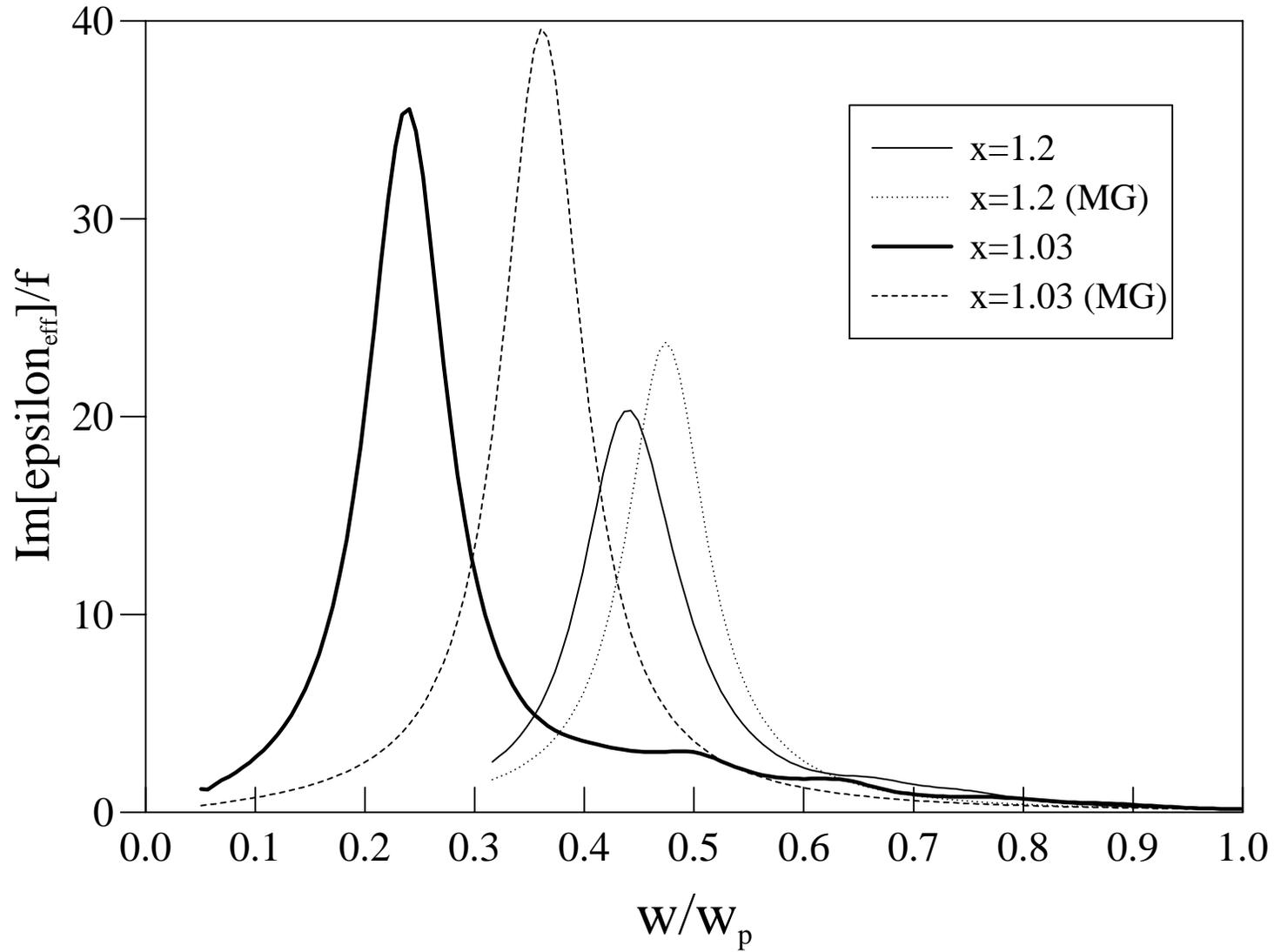
\begin{figure}
\caption{The same as in Fig. 2, for filling fractions of $54.5\%$ and $74.0\%$. 
Full thin and thick lines: our numerical results. Dotted and dashed lines: MG results.}
\end{figure}

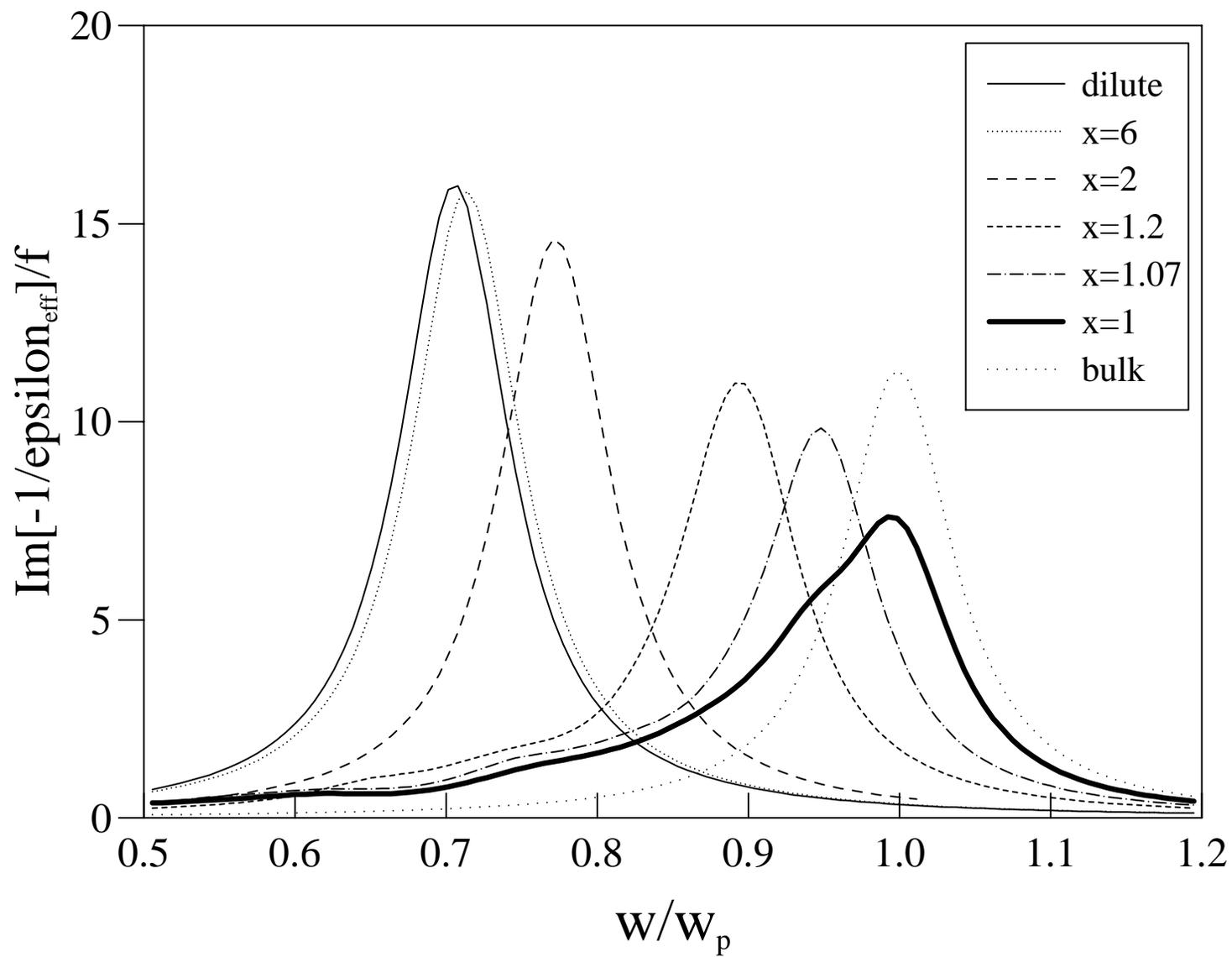
\begin{figure}
\caption{The effective energy-loss function of the periodic system described in Fig.
1, for $p$ polarized electromagnetic waves. Full thin line: the isolated cylinder result
($f\to 0$). Dotted, long-dashed, short-dashed, dashed-dotted and thick full lines: our
numerical results for volume filling fractions of $2.2\%$, $19.6\%$, $54.5\%$, $68.6\%$
and $78.5\%$, respectively. The bulk energy-loss function is represented bay a dotted
line.}
\end{figure}

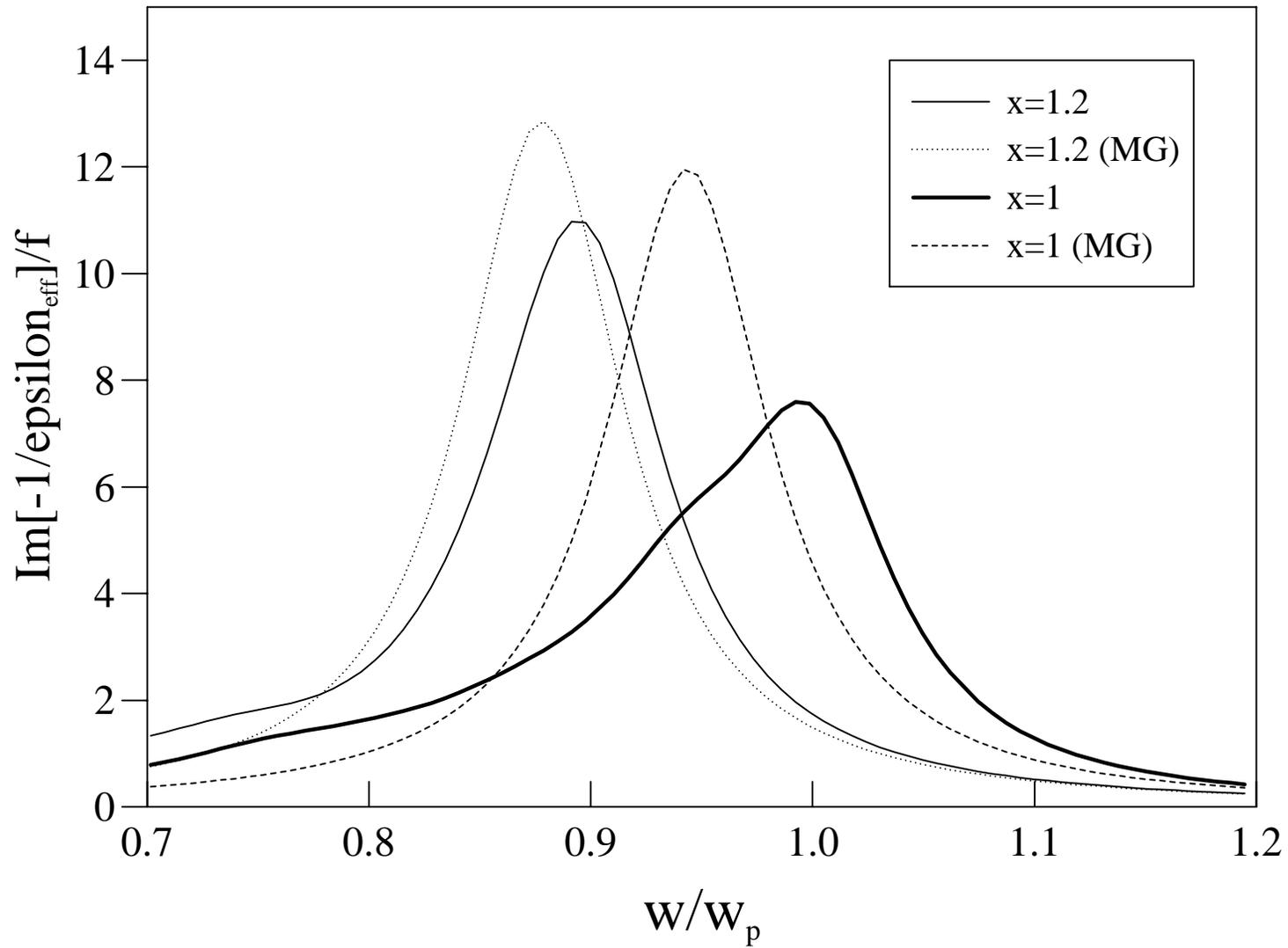
\begin{figure}
\caption{The same as in Fig. 4, for filling fractions of $54.5\%$ and $78.5\%$. 
Full thin and thick lines: our numerical results. Dotted and dashed lines: MG results.}
\end{figure}

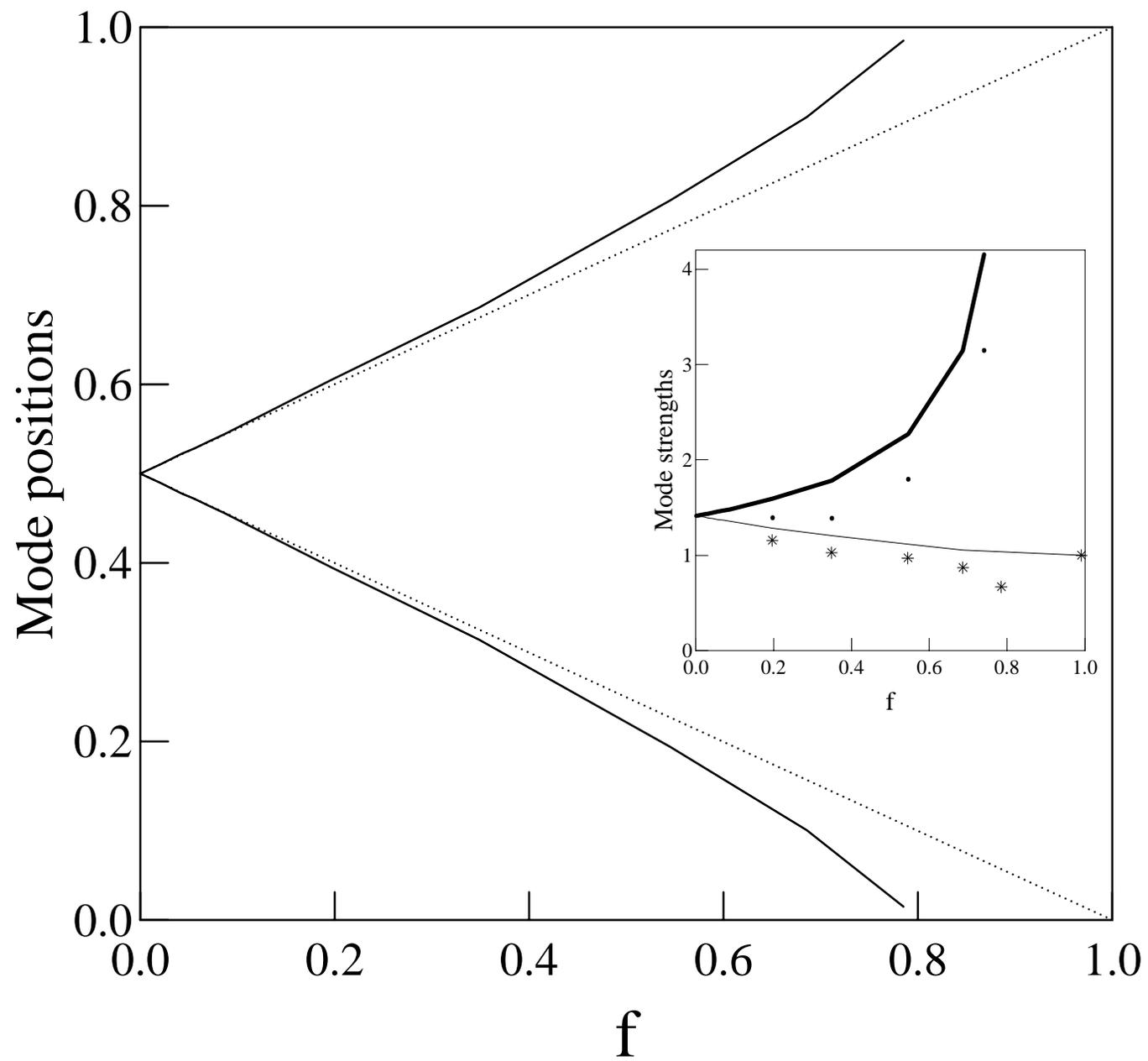
\begin{figure}
\caption{Dipolar mode positions (depolarization factors), as a function of the volume
filling fraction
$f$. Solid lines: our numerical results for $m_1$ (the line below the dilute limit of
$1/2$) and
$n_1$ (the line above the dilute limit of $1/2$). Dotted lines represent the MG
depolarization factors given by Eqs. (10) and (11). Dipolar mode strengths are
represented in the inset. Dots and stars: our numerical results for the dipolar peak
heights in
${\rm Im}\left[\epsilon_{\rm eff}(\omega)\right]/(f\,H)$ and
${\rm Im}\left[-\epsilon^{-1}_{\rm eff}(\omega)\right]/(f\,H)$, respectively. Full thick
and thin lines:
$1/\protect\sqrt\protect m_1$ and $1/\protect\sqrt\protect n_1$, respectively. Here,
$H=\omega_p/\gamma$.}
\end{figure}

\end{document}